\documentclass[dvipdfmx]{PoS}

\title{Nucleon isovector couplings from 2+1 flavor lattice QCD at the physical point}

\ShortTitle{Nucleon isovector couplings from 2+1 flavor lattice QCD at the physical point}

\author{\speaker{Natsuki Tsukamoto$^a$}, Yasumichi Aoki$^b$, Ken-Ichi Ishikawa$^c$,  Yoshinobu Kuramashi$^{d}$, Eigo Shintani$^b$, Shoichi Sasaki$^a$, Takeshi Yamazaki$^{e.d}$\\
        ${}^a$Department of Physics, Tohoku University, Sendai 980-8578, Japan \\
	${}^b$RIKEN Center for Computational Science, Kobe 650-0047, Japan \\
	${}^c$Graduate School of Science, Hiroshima University, Higashi-Hiroshima 739-8526, Japan \\
	${}^d$Center for Computational Sciences, University of Tsukuba, Tsukuba 305-8577, Japan \\
	${}^e$Faculty of Pure and Applied Sciences, University of Tsukuba, Tsukuba 305-8571, Japan
\\
\\        E-mail: \email{tsukamoto@nucl.phys.tohoku.ac.jp}
}

\abstract{We present results on the axial, scalar and tensor isovector-couplings of the nucleon from 2+1 flavor lattice QCD with physical light quarks ($m_\pi$ = 135~MeV) in large spatial 
volume of (10.8~fm)$^3$. 
The calculations are carried out with the PACS10 gauge configurations generated by the 
PACS Collaboration 
with the stout-smeared $\mathcal{O}(a)$ improved Wilson fermions and Iwasaki gauge action 
at $\beta=1.82$ corresponding to the lattice spacing of 0.084~fm. 
For the renormalization, we use the RI/SMOM scheme, a variant of Rome-Southampton RI/MOM scheme with reduced systematic errors, as the intermediate scheme.
We then evaluate our final results in the $\overline{\rm MS}$ scheme at a scale of 2~GeV, using the continuum perturbation theory for the matching scale of RI/SMOM and $\overline{\rm MS}$ schemes and running.
}

\FullConference{37th International Symposium on Lattice Field Theory - Lattice2019\\
		16-22 June 2019\\
		Wuhan, China}

\usepackage{amsmath}	
\begin{document}

\section{Introduction}

Future and current precision {$\beta$}-decay measurements with cold and ultracold neutrons 
provide us an opportunity to study the sensitivity of the nucleon isovector matrix elements to 
new physics beyond the standard model (BSM). The neutron life-time puzzle associated with 
the nucleon axial coupling ($g_A$) is one of such examples~\cite{Czarnecki2018}. The nucleon scalar and tensor 
couplings ($g_S$ and $g_T$) should play important roles to constrain the limit of 
non-standard interactions mediated by undiscovered gauge bosons in the scalar and tensor channels
if the BSM contributions are present~\cite{Bhattacharya2012}. Especially the nucleon scalar isovector-coupling, 
which is related to the mass difference between the light quarks, is a phenomenologically 
interesting quantity~\cite{Gonzalez-Alonso2014}.
On the other hand, the tensor coupling has the same transformation properties under $P$ and $T$ discrete symmetries as the electric dipole moment (EDM) current. Thus the nucleon tensor 
isovector-coupling is also an important information regarding the size of neutron EDM.

Although the vector and axial isovector-couplings ($g_V$ and $g_A$) are 
well measured in both experiment and lattice QCD, the scalar and tensor isovector-couplings 
are so far not accessible in experiment.
On the other hand, lattice determination of the scalar and tensor isovector-couplings have recently 
performed by several groups~\cite{Yoon2017,Hasan2019,Alexandrou2017}. 
Further comprehensive studies of the nucleon isovector-couplings including $g_S$ and  $g_T$ 
as well as $g_V$ and $g_A$ are still needed.

\section{Method}
In general the isovector nucleon couplings $g_O$ are expressed by 
the neutron-proton transition matrix element with the quark charged (off-diagonal) currents
\begin{align}
\langle p(p,s)|\bar{u}\Gamma^{O} d|n(p,s)\rangle = g_O \overline{u}_p(p,s)
\Gamma^O u_n(p,s),
\end{align}
where $\Gamma^O$ is a Dirac matrix appropriate for the channel $O$
($O=V, A, S, P, T$).
Considering the $SU(2)$ Lie algebra associated with isospin,
the isovector nucleon matrix element can be rewritten by the proton matrix element of the diagonal isospin current 
%
%
\begin{align}
   \langle p | \overline{u}\Gamma^{O} d | n\rangle =  \langle p | \overline{u}\Gamma^{O} u | p\rangle
   - \langle p | \overline{d}\Gamma^{O} d | p\rangle
\end{align}
in the isospin limit~\cite{Sasaki:2003jh}. Therefore, the isovector couplings  are related with the flavor diagonal couplings $g^{f}_O =\langle p | \overline{f}\Gamma^O f | p\rangle$ with $f=u$ or $d$ as 
%
%
$g_O^{\rm isovector} = g^{u}_{O} - g^d_{O}$.
%

In order to calculate the nucleon matrix element in lattice QCD simulations, we compute the three-point correlation functions consisting of the smeared proton source and sink operators ($N$ and $\overline{N}$) with a given bilinear operator $\mathcal{J}^O=\overline{u}\Gamma^O u - \overline{d}\Gamma^O d$
\begin{align}
C_{O}^{\mathcal{P}}(t, \mathbf{p}^\prime, \mathbf{p}) = \frac{1}{4}{\rm Tr}\left\{
\mathcal{P}\langle
 N(t_{\rm sink},\mathbf{p}^\prime)\mathcal{J}^O(t, \mathbf{q})\overline{N}(t_{\rm src}, -\mathbf{p})
   \rangle \right\},
\end{align}
where $\mathbf{q}=\mathbf{p}-\mathbf{p}^\prime$ represents the three dimensional
momentum transfer. 
A well-known procedure for determining the couplings is to calculate the following
ratio fo the three-point and two-point correlation functions with zero momentum transfer
\begin{align}
\frac{C_{O}^{\mathcal{P}}(t, \mathbf{0}, \mathbf{0})}{C_{\rm 2pt}(t_{\rm sink} - t_{\rm  src})} &\rightarrow g_O^{\rm bare} & {\rm for} \ \ \ \ \  t_{\rm sink} \gg t \gg t_{\rm src}
\end{align}
where $C_{\rm 2pt}(t_{\rm sink} - t_{\rm  src})$ represents the proton two-point correlation function 
with the same smeared source and sink at the rest frame. Recall that the ratio vanishes unless $\Gamma^O=1(S)$, $\gamma_4(V)$, $\gamma_i\gamma_5(A)$, and $\sigma_{ij}(T)$ with $i,j=1,2,3$~\cite{Sasaki:2003jh}.
The nonvanishing ratio gives an asymptotic plateau corresponding to the bare value of the
coupling $g_O$ relevant for the $O$ channel. In this study we focus on the axial ($A$), scalar ($S$) and 
tensor ($T$) couplings.

\section{Simulation Details}
We mainly used the PACS10 configurations~\cite{Ishikawa2018} generated by the PACS Collaboration with the stout-smeared $\mathcal{O}(a)$ improved Wilson-clover fermions and Iwasaki gauge action.
Two lattice sizes are used for this study, $128^4$ and $64^4$, corresponding to linear spatial extents of approximately 10 and 5 fm (See also Tab.~\ref{tab:sim}). The smaller volume ensembles are used only for computing the renormalization constant which is known to be less sensitive to the finite volume effects, while our main results of the nucleon matrix elements are obtained from the larger volume ensembles.
The simulation details are summarized in Tab.~\ref{tab:sim}.

\begin{table}[h]
 \begin{center}
\begin{tabular}{ccccc}\hline
  $L^3\times T$ & $a^{-1}$[GeV] & $\kappa_s$  & $\kappa_l$ & $M_\pi$ [GeV]  \\\hline
 $128^3\times 128$ & 2.3 &  0.124902 & 0.126117 & 0.135~\cite{Shintani2018} \\
$64^3 \times 64$ & 2.3 &  0.124902 & 0.126117 & 0.139~\cite{Ishikawa2018} \\
\hline
\end{tabular} 
\caption{Simulation Details\label{tab:sim}}
   \end{center}
 \end{table}

\section{Preliminary Results}
\subsection{Updates from the previous results}
In our previous study~\cite{Shintani2018}, we had computed nucleon two-point and three-point correlation functions using the exponentially smeared source and sink with four different source-sink 
separations ($t_{\rm sep}=t_{\rm sink}-t_{\rm src}$). 
Significant reduction of the computational cost is achieved by employing the all-mode-averaging (AMA) method optimized by the delation technique~\cite{Shintani2015}.
We then obtained five basic quantities of the nucleon from nucleon form factors: 
the electric and magnetic root-mean-square (RMS) radii, the magnetic moment, the axial isovector-coupling ($g_A$), and the axial RMS radius, with good statistical precisions with within 2-5\%. It is worth mentioning that the 2\% precision of $g_A$ whose value is fairly consistent with the experimental one, was achieved~\cite{Shintani2018}. At present we are pursuing one percent-level precision on $g_A$. Meanwhile, we focus on an accurate
determination of the scalar and tensor isovector-couplings ($g_S$ and $g_T$).

In this study, the nucleon three-point correlation functions are calculated using the sequential
source method with a fixed source~\cite{Sasaki:2003jh}. We adopt the source-sink separation of 
$t_{\rm sep}/ a = 13$ and 16 with the gauge-covariant Gauss-smeared source and sink. 
The number of measurements used in this study is listed in Tab.~\ref{tab:nofmeas} together
with our previous study performed with the exponentially smeared operators.

\begin{figure}
 \includegraphics[width=0.48\textwidth]{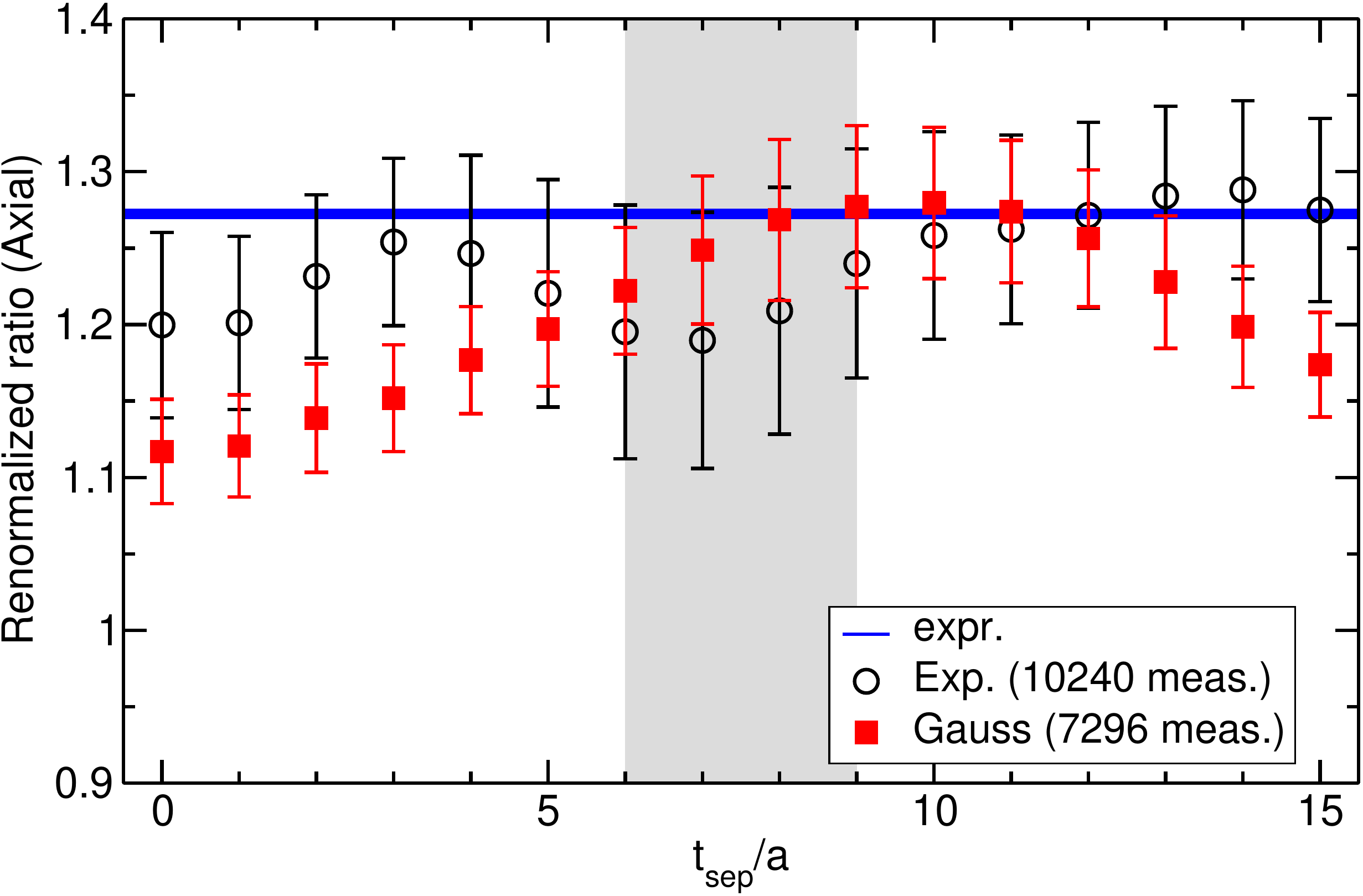} 
 \includegraphics[width=0.48\textwidth]{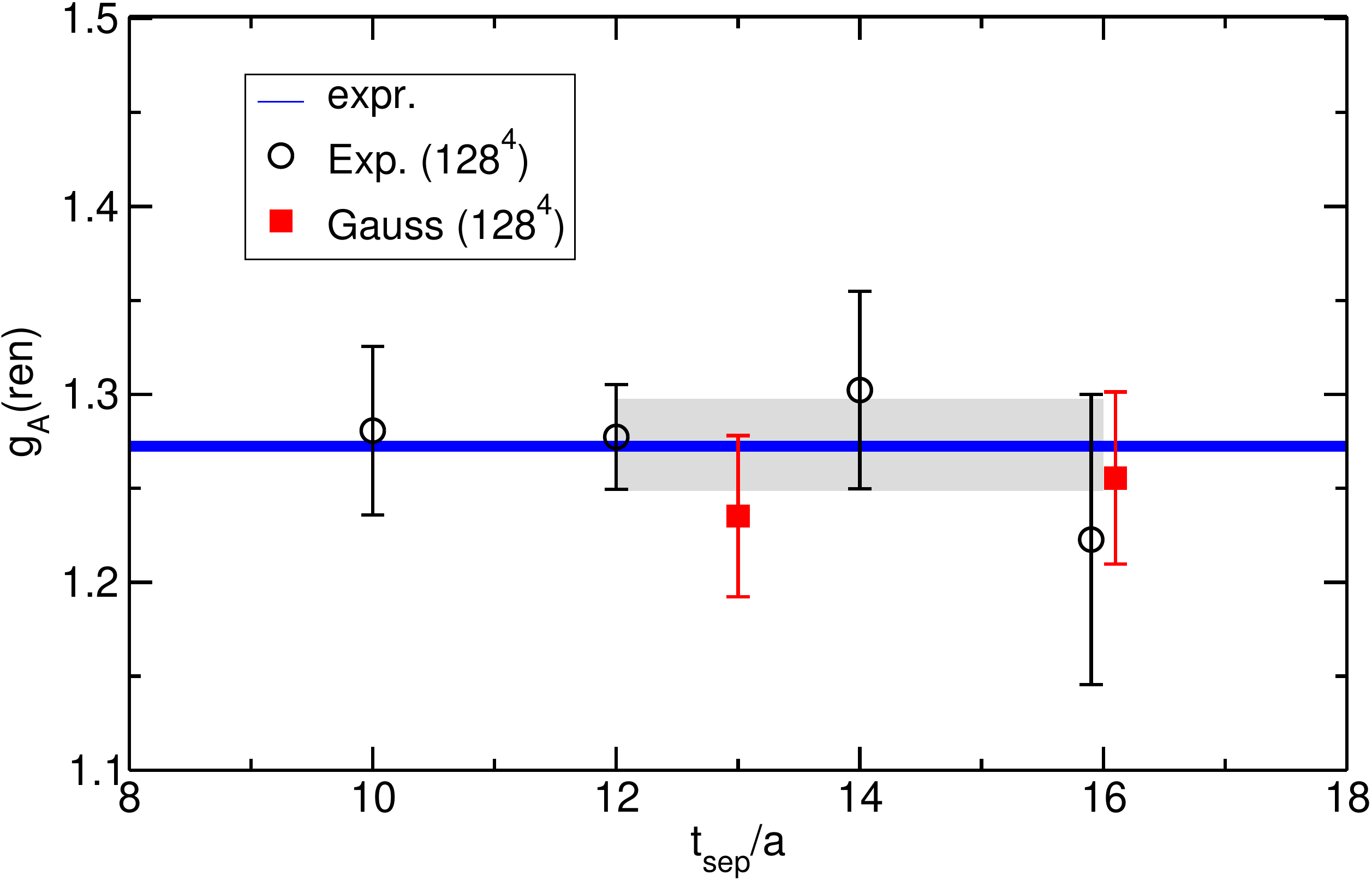} 
\caption{Comparing the results obtained from two types of the smearing: The black circles and blue squares present results obtained from the exponential and Gaussian smearings respectively.
The gray band in the right panel on the figure represents the combined value with $t_{\rm sep}/a =\{12, 14, 16\}$ using the exponentially smeared operators, that was quoted in Ref.~\cite{Shintani2018}.
\label{fig:EXPvsGAUSS}}
\end{figure}

In the left panel of Fig.~\ref{fig:EXPvsGAUSS}, we plot
the ratio of the relevant three-point and two-point correlation functions for the axial channel with $t_{\rm sep}/a=16$ as a function of the current insertion time $t$ as a typical example. 
The $t$-dependence of the ratio is mild in both smearing types 
of the exponential (Exp.) and Gaussian (Gauss) forms.
The local axial current is renormalized with the value of $Z_A=0.9650(68)(95)$ obtained with the Schr\"odinger functional (SF) scheme~\cite{Ishikawa2015}.
As shown in Fig.~\ref{fig:EXPvsGAUSS}, the statistical uncertainties on results from the Gauss smeared operators are almost twice smaller than that of the exponentially smeared operators at $6 \le t/a \le 9$ (gray shaded band), though the total number of the former measurements are about 1.5 times smaller than the latter.  
We found that for the same statistical accuracy, the total computational cost of the Gauss smeared 
operators is roughly 5-6 times lower than the case of the exponentially smeared operator.

We next show the $t_{\rm sep}$ dependence of the renormalized axial coupling in both cases of 
the exponential and Gaussian smearings in the right panel of Fig.~\ref{fig:EXPvsGAUSS}. 
As one can easily see, when the Gauss smeared operators are adopted, more precise determination of $g_A$ is achieved even with $t_{\rm sep}/a =16$, which is a maximum size of sink-source separation in our previous work. This figure shows that our results of the renormalized $g_A$ 
in all cases of $t_{\rm sep}$ agree with the experimental value, 1.2724(23) (denoted as a blue line). We do not observe a significant $t_{\rm sep}$ dependence.
We also expect that our final result of $g_A$ could eventually reach {\it one percent-level precision}
from the combined value with $t_{\rm sep}/a =\{13, 16\}$ even {\it at the physical point}.

\begin{table}
\begin{center}
 \begin{tabular}{|cccc|cccc|}\hline
  Smearing-type & $t_{\rm sep}/a$ &$N_{\rm conf}$  & \# of meas. &   Smearing-type & $t_{\rm sep}/a$& $N_{\rm conf}$ & \# of meas. \\ \hline
Exp. &  10 & 20 & 2560 & Gauss & 13 & 16 & 1024 \\
 &12 & 20 & 5120 &  & 16 & 19 & 7296 \\
 &14 & 20 &6400 &  & &  & \\ 
 &16 & 20 &10240 &  &  &  & \\ 
\hline
 \end{tabular}
\caption{The total number of measurements at each source-sink separation
with two smearing types.\label{tab:nofmeas}}
\end{center}
\end{table}

\subsection{Scalar and tensor couplings}

\begin{figure}
 \includegraphics[width=0.48\textwidth]{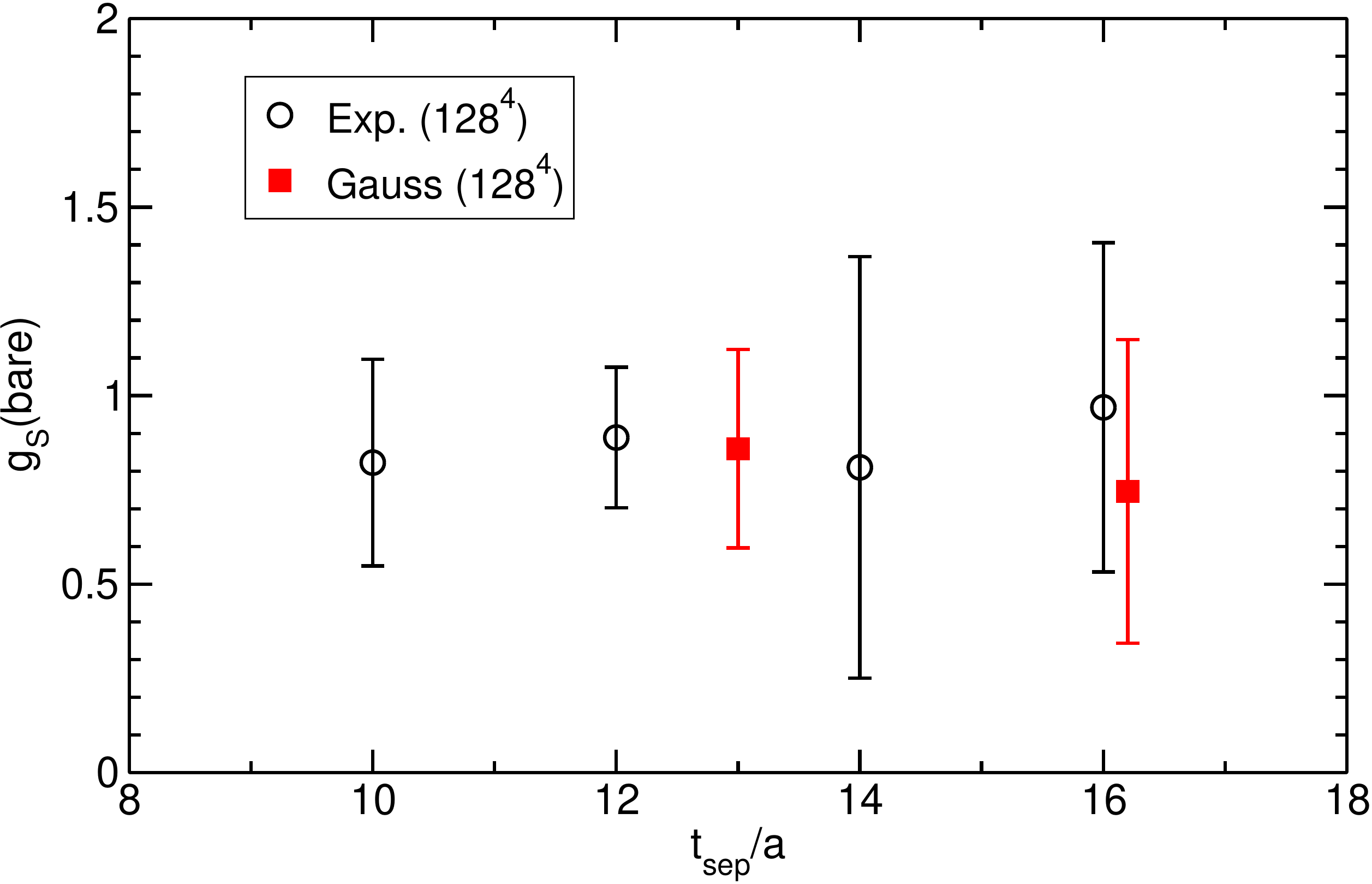}
 \includegraphics[width=0.48\textwidth]{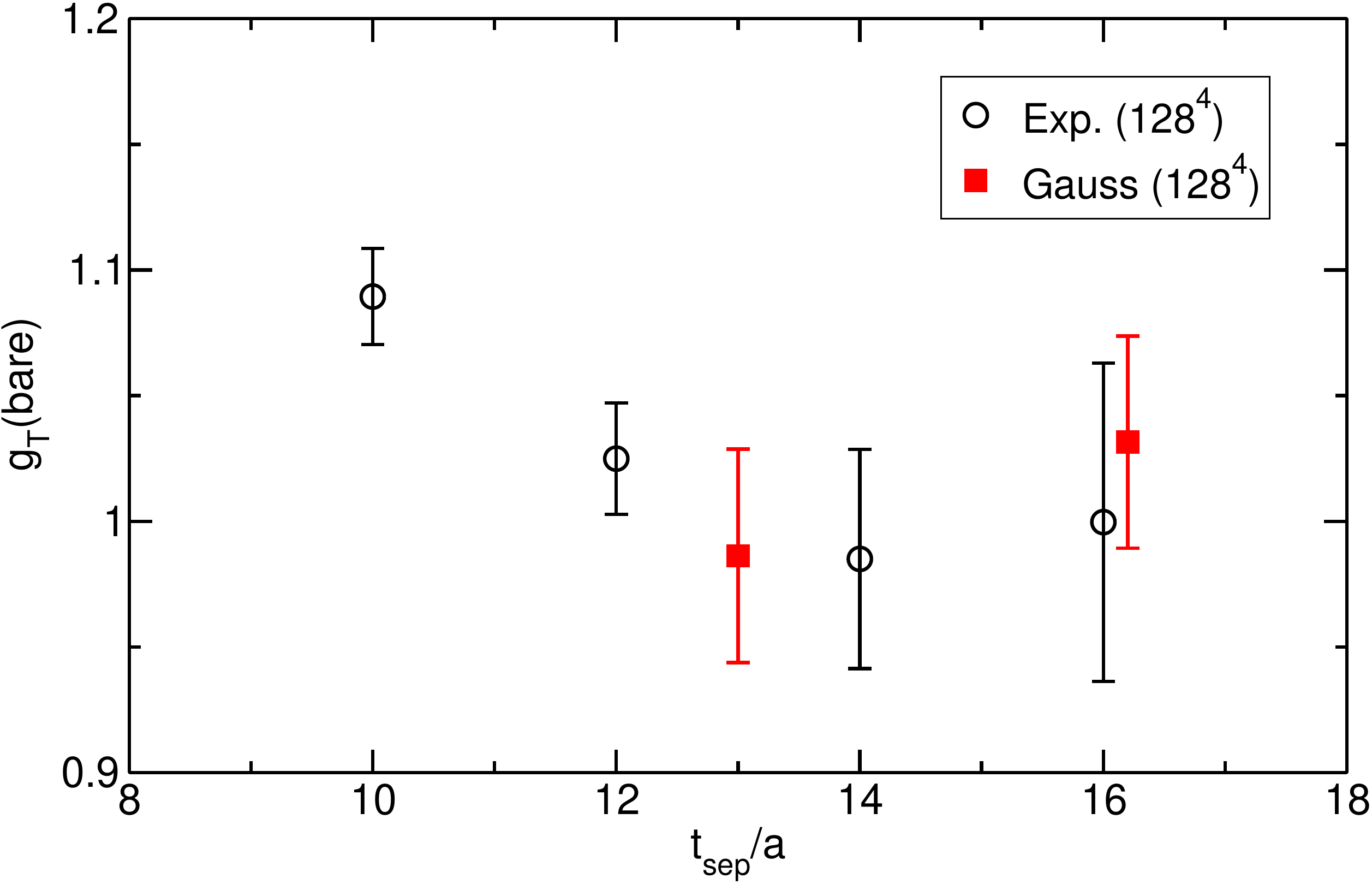}
\caption{
Bare couplings for the scalar (left) and tensor (right) channels as a function of $t_{\rm sep}$.\label{fig:barecoup}
}
\end{figure}

In Fig.~\ref{fig:barecoup}, we show the results for the bare values of $g_S$ and $g_T$, which are 
obtained with several different source-sink separations. The $t_{\rm sep}$ dependence appears slightly for the case of $t_{\rm sep}/a <12$ in the tensor channel, while there is 
no $t_{\rm sep}$-dependence
in the scalar channel albeit with rather large statistical errors. 
The systematic uncertainties stemming from the excited state
contamination are enough small for the cases of $t_{\rm sep}/a \geq 13$ within the statistical errors.

In order to be compared with the experiment values or other lattice results, the bare couplings
$g_S$ and $g_T$ should be renormalized with the renormalization factors $Z_S$ and $Z_T$ in the certain scheme. As for the scalar and tensor channels, we first renormalize the scalar and tensor
couplings nonperturbatively using the Rome-Southampton method,
where the regularization independent momentum subtraction scheme is adopted.
The renormalization factors determined in the RI/MOM subtraction scheme are converted 
to the $\overline{\rm MS}$ scheme and then 
evolved to the scale of 2 GeV using the perturbation theory.

The following procedure is performed to evaluate the renormalization factors in this study:
\begin{enumerate}
\item Obtain the renormalization factors $Z_S/Z_{V(A)}$ and $Z_T/Z_{V(A)}$ in the RI/SMOM scheme~\cite{Sturm2009} at certain scale of $\mu_0$. Using the value of $Z_V$ ($Z_A$) obtained in the SF scheme~\cite{Ishikawa2015}, we can determine the renormalization factors $Z^{RI}_O(\mu_0)$ ($O=S$ and $T$)
in the fully nonperturbative manner.

 \item Convert $Z^{RI}_O$ into the $\overline{\rm MS}$ scheme at the matching scale 
 $\mu_0$ and then evolve the renormalization factors $Z_O^{\overline{\rm MS}}(\mu_0)$ to the scale of 2 GeV with a help of the continuum two-loop perturbation theory~\cite{Almeida2010}. Here, in principle, $Z_O^{\overline{\rm MS}}(\mu_0; 2~{\rm GeV})$ are supposed to be insensitive to the choice of the matching scale $\mu_0$ within a certain range.
 
 \item Eliminate the residual dependence on the choice of the matching scale $\mu_0$
 appearing in the value of $Z_O^{\overline{\rm MS}}(\mu_0; 2~{\rm GeV})$ due to the presence of
 lattice artifacts at higher $\mu_0$ and truncation of the perturbative series at lower $\mu_0$.

 \end{enumerate}

\begin{figure}
 \includegraphics[width=0.48\textwidth]{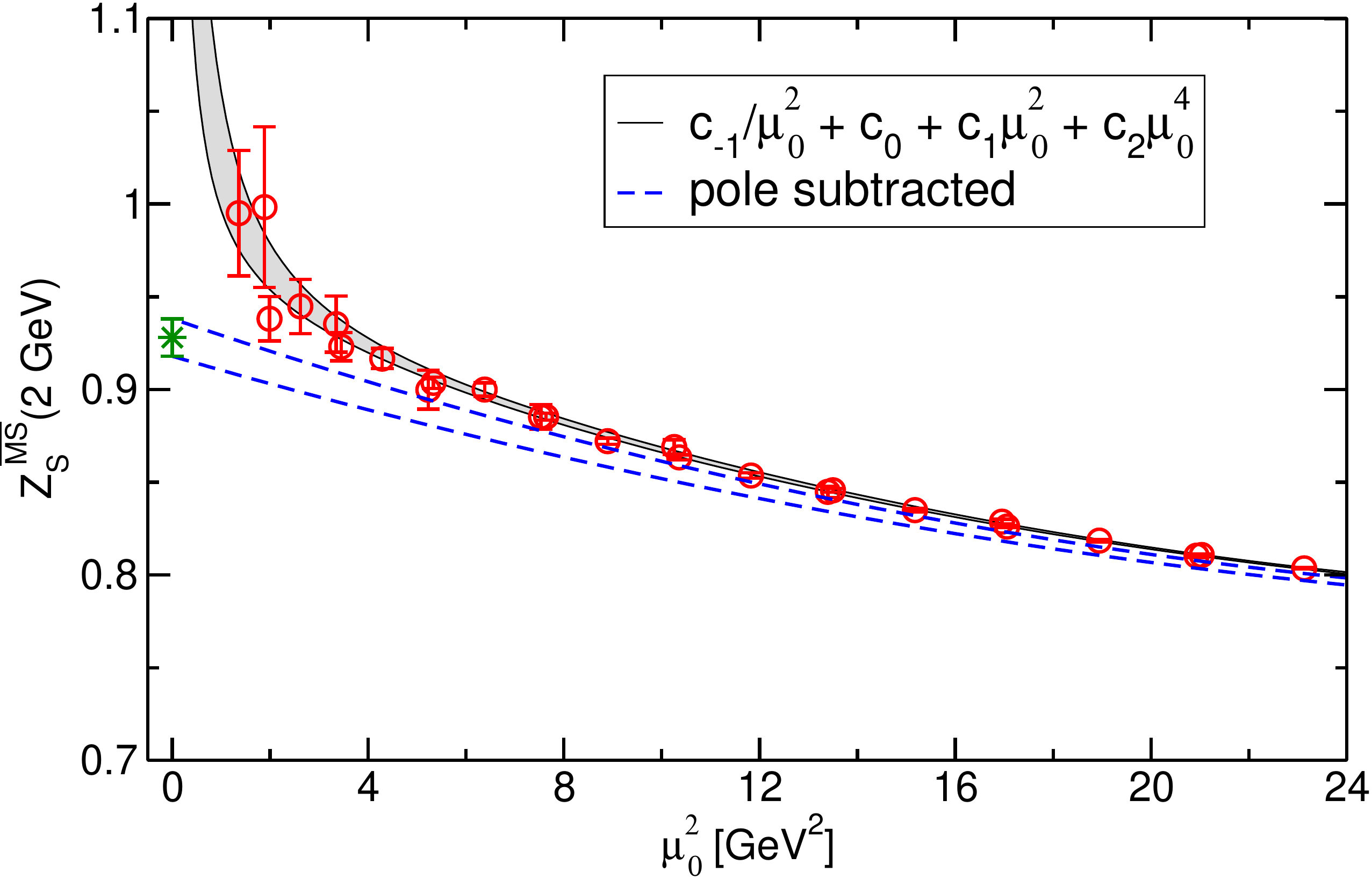}
 \includegraphics[width=0.48\textwidth]{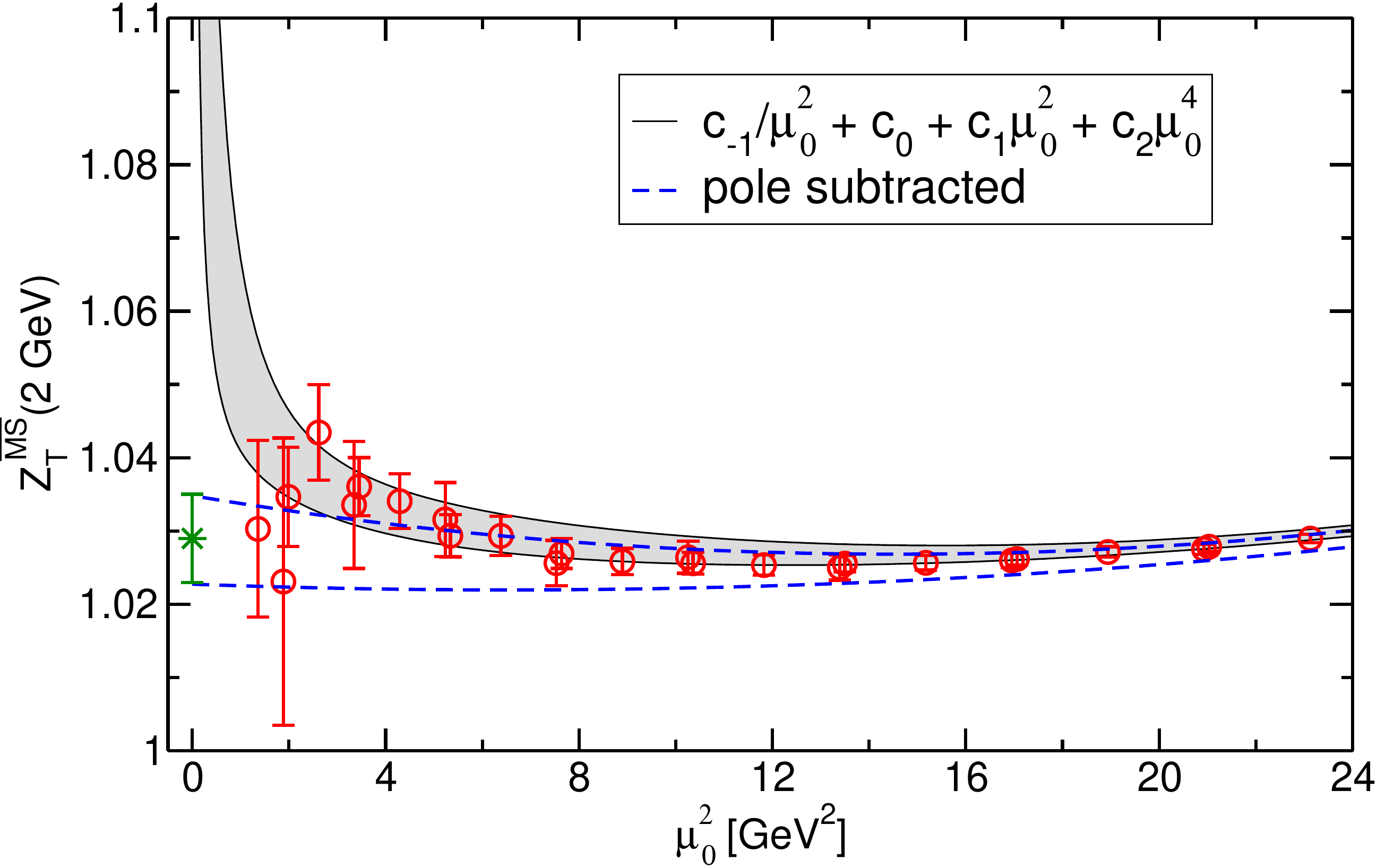}
\caption{The matching scale dependence of  $Z_S^{\overline{\rm MS}}(2~{\rm GeV})$
(left) and $Z_T^{\overline{\rm MS}}(2~{\rm GeV})$ (right).\label{fig:renorm}}
\end{figure}

In Fig.~\ref{fig:renorm}, we show the value of $Z_O^{\overline{\rm MS}}(\mu_0; 2~{\rm GeV})$
for the scalar (left panel) and tensor (right panel) as a function of $\mu_0^2$. The residual
dependence of the choice of the matching scale $\mu_0$ appears more largely in the scalar
channel than the tensor channel where the residual dependence is not significant.
In order to eliminate the residual $\mu_0$ dependence, we used the following functional form~\cite{Yoon2017,Hasan2019}:
\begin{align}
 Z_O^{\overline{\rm MS}}(\mu_0; 2~{\rm GeV}) = \frac{c_{-1}}{\mu_0^2} + c_0 + c_1 \mu_0^2 + c_2 {\mu_0}^4
 \label{Eq:fitform}
\end{align}
with $c_0$ being the $\mu_0$-independent value of $Z_O^{\overline{\rm MS}}(2~{\rm GeV})$.
A pole term in Eq.~(\ref{Eq:fitform}) should be originated from the existence of dimension two condensate in the Landau gauge as the nonperturbative effect~\cite{Boucaud2006}.
The fit results with the form~(\ref{Eq:fitform}) are represented by
gray shaded curves in Fig.~\ref{fig:renorm}. Blue dashed curves are given
after the pole contribution is subtracted. The constant term $c_0$ can be read off
from the blue dashed curve as y-axis intercept in each panel.

\begin{figure}
 \includegraphics[width=0.48\textwidth]{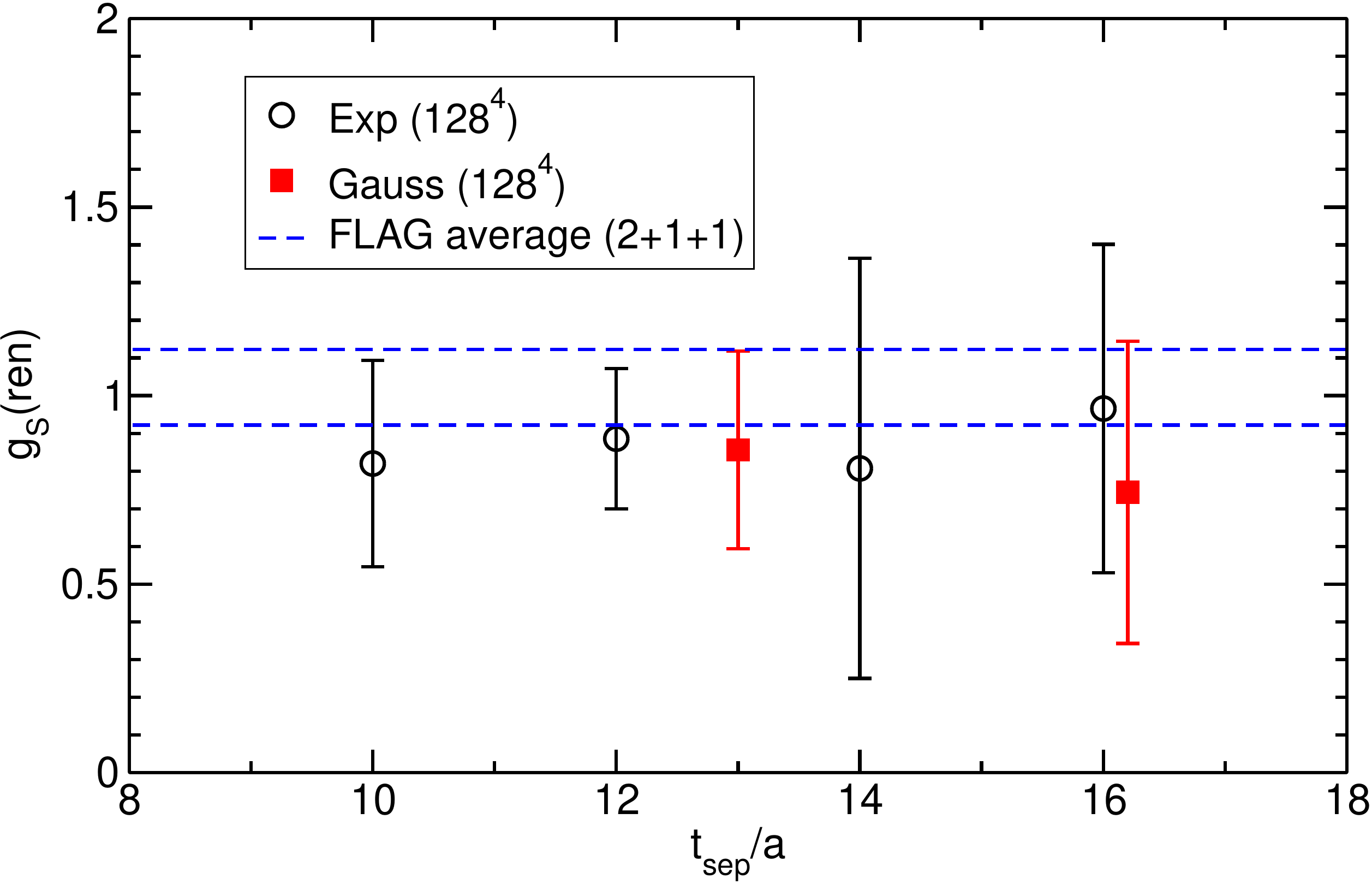}
 \includegraphics[width=0.48\textwidth]{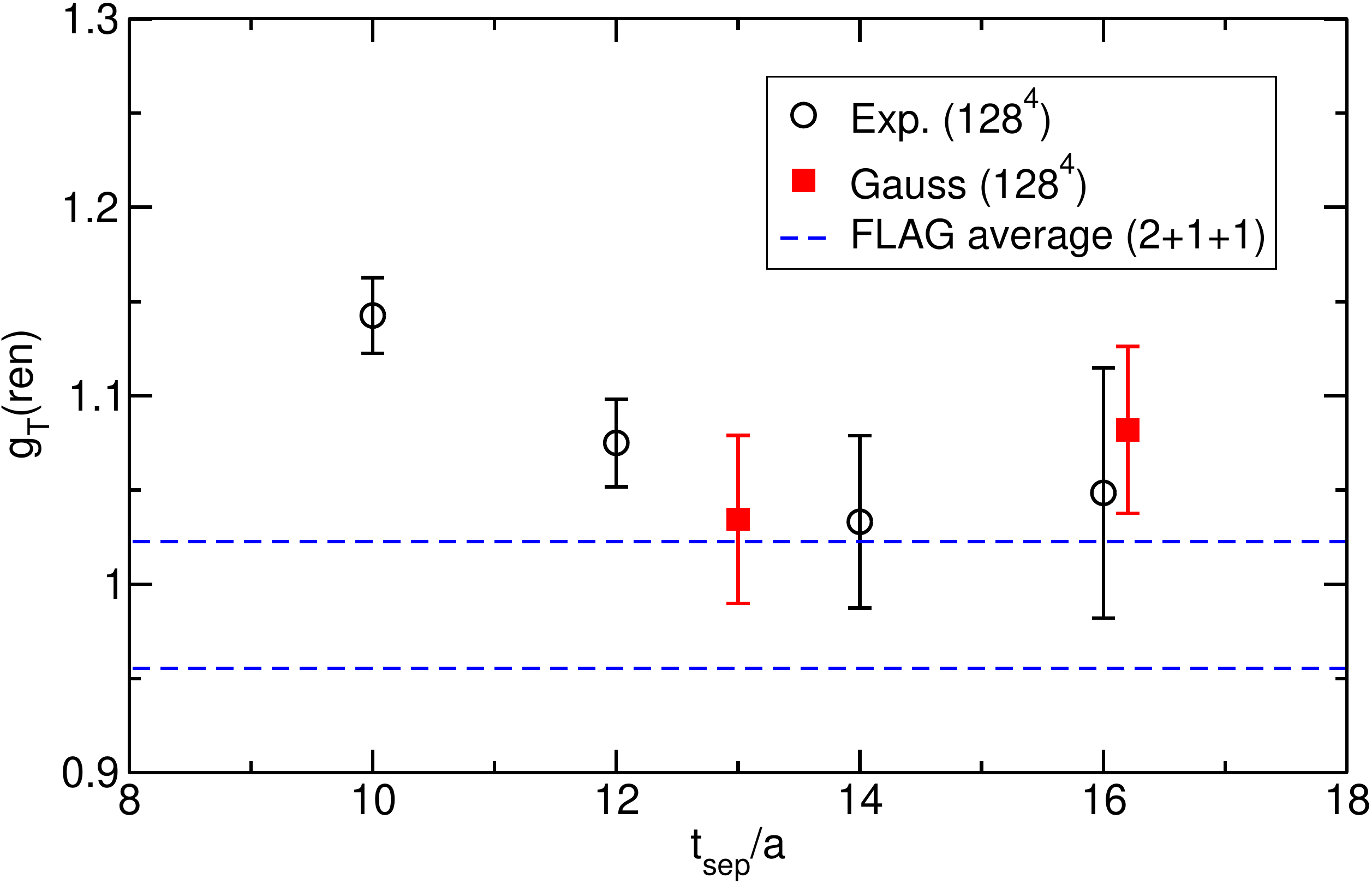}
\caption{
Renormalized scalar (left) and tensor (right) couplings in the $\overline{\rm MS}$ scheme at the scale of 2~GeV. \label{fig:ren_coup}}
\end{figure}

Combining the renormalization factors with the bare couplings, 
we finally obtain the renormalized values of $g_S$ and $g_T$ in the $\overline{\rm MS}$
scheme at the scale of 2 GeV, which are consistent with the FLAG
average values~\cite{Aoki2019} as shown in the left and right panels of Fig.~\ref{fig:ren_coup} 
for $g_S$ and $g_T$ respectively.  

\section{Summary}
We have calculated the axial, scalar and tensor isovector-couplings of the nucleon
using the PACS10 gauge configurations. To improve the statistical accuracy and
reduce the computation time from our previous study, we use the Gauss-smeared 
operators for calculating the relevant three-point and two-point correlation functions
of the proton. We found that the Gauss-smeared operators efficiently reduce
the statistical uncertainties on the ratio of the three-point and two-point correlation functions
in comparison to the exponentially smeared operators adopted in our previous study.
Indeed, the same level of precision on the determination of $g_A$ with the large source-sink separation  
$t_{\rm sep}/a = 16$ was easily achieved with 
the roughly 5-6 times lower computational cost than the previous calculations.
Combining the results with $t_{\rm sep}/a = 13, 16$, we can expect that 
the usage of the Gauss-smeared operator enables to us to reach one percent-level precision 
for our final result of $g_A$ .

We also calculated the isovector couplings in the scalar and tensor channels with different source-sink separations and found that the systematic uncertainties stemming from the excited state contamination are 
well under control for $g_S$ and $g_T$ as well as $g_A$ in our study.
We also nonperturbatively estimated the renormalization factors for
the scalar and tensor current using the RI/SMOM scheme. We finally determine the renormalized value of $g_S$ and $g_T$ in conversion to the $\overline{\rm MS}$ scheme. Our results are consistent with the FLAG average values~\cite{Aoki2019}.

\section*{Acknowledgement}
N.~Tsukamoto is supported in part by the Joint Research Program of 
RIKEN Center for Computational Science 
(R-CCS).
The present research used computational resources 
through the HPCI System Research Project (hp140155, 
hp150135, 
hp160125, 
hp170022, 
hp180051, 
hp180072, \\
hp180126, 
hp190025, 
hp190055). 
This work is supported in part by 
Grant-in-Aid for Scientific Research (Nos. 16K05320, 16H06002, 18K03605, 19H01892) and the U.S.-Japan Science and Technology Cooperation Program in High Energy Physics for FY2018.

\bibliographystyle{h-physrev4}

\bibliography{./Proc.bib}

\end{document}